\documentstyle{europhys}

\def\And{{\rm and\ }}

\def\stars{\bigskip\centerline{***}\medskip}

\newif\ifboo \boofalse

\def\Review#1{\boofalse{\it #1},}
\def\Name#1{{\sc #1},}
\def\Vol#1{\ifboo Vol. {\bf #1}\else{\bf #1}\fi}
\def\Year#1{\ifboo #1\else(#1)\fi}
\def\Book#1{\bootrue{\it #1},}
\def\Page#1{\ifboo {\rm p. #1}\else{\rm #1}\fi}

\begin{document}

\euro{}{}{}{}
\Date{}
\title{Ratchet Potential for Fluxons in Josephson-junction Arrays}

\author{F. Falo\inst{1,}, P.J. Mart\'{\i}nez\inst{2}, J.J.
Mazo\inst{1,}\inst{3}, S. Cilla\inst{1,}
}
\institute{
    \inst{1} Dep. de F\'{\i}sica de la Materia Condensada e Instituto de 
    Ciencia de Materiales de Arag\'on C.S.I.C. - Universidad de Zaragoza, 50009 Zaragoza, Spain \\
    \inst{2} Dep. de F\'{\i}sica Aplicada e Instituto de 
    Ciencia de Materiales de Arag\'on C.S.I.C. - Universidad de Zaragoza, 50009 Zaragoza, Spain \\
    \inst{3} Dep. of Electrical Engineering and Computer Science,
Massachusetts Institute of Technology, Cambridge, Massachusetts 02139}
\rec{}{}
\pacs{
\Pacs{74}{50$+$r}{Proximity effects, weak links, tunneling phenomena and
Josephson  effects}
\Pacs{05}{40$+$j}{Fluctuation phenomena, random processes and Brownian motion.}
\Pacs{85}{25$+$Na}{Superconducting microelectronic circuits.}
}
\maketitle
\begin{abstract}
We propose a simple configuration of a one-dimensional parallel array of
Josephson-junctions in which the pinning potential for trapped fluxons
lacks inversion symmetry (ratchet potential). This system can be modelised
by a set of non-linear pendula with alternating lenghts and asymmetric
harmonic couplings. We show, by molecular
dynamics simulation, that fluxons behave as single particles in which the
predictions for overdamped thermal ratchets can be easily verified.
\end{abstract}

Directional motion (DM) of brownian particles has been a target 
of basic and
applied research in the last five years. 
The initial motivation and interest in this field came from cell biology:
The study of the mechanism of vesicles transport inside eukariotic cells,
via motor proteins along microtubules \cite{RMP,Magnasco}. 
Later on new systems with
the same underlying ideas for transportation were proposed. Those systems
include: phase separation engines\cite{Astumian}, drop motion under 
ac forces\cite{drops}, growth of
surfaces\cite{Growth}, and rectification in asymmetric superconducting rings 
(SQUID) \cite{Sols}.
To have DM of a particle submitted to a periodic one-dimensional
potential 
$V(x)=V(x+L)$ it is needed
i) broken spatial symmetry i.e. $V(x) \neq V(-x)$ and
ii) to drive the particle out of thermal equilibrium. This last 
condition makes possible to extract work from the system.

The first condition is usually fulfilled by the use of an ``ad hoc''
ratchet
potential \cite{Magnasco}. The latter condition can be implemented
in different ways:
by an oscillating driving field \cite{Chialvo} , 
by a time-correlated non-thermal noise  \cite{Magnasco}
or a fluctuating $V(x)$ potential \cite{fluctu}, among others. 
In all these cases, the conditions for the fluctuation-dissipation 
theorem do not hold.

In this paper we propose a {\em new experimental realization of 
DM in a very simple and controllable system}: a parallel
Josephson-junction array. Such system has become an excellent realization of
the Frenkel-Kontorova (FK) or discrete sine-Gordon model \cite{Watanabe}.
Fluxons (or kinks) in the array move along a periodic pinning 
potential (the so called Peierls-Nabarro potential). We will show that in a 
simple geometry this potential is a ratchet one and a fluxon 
behaves as a particle with well defined directional motion.

This Letter is organized as follows: we begin by describing the basic assumptions 
for the array and the equations of motion for the phases
are obtained. Then, we calculate the energy profile for a fluxon trapped in the
array and show its ``ratchet'' character. Finally, we compare our results 
for the dynamics of the fluxon 
with what is expected from single particle dynamics.

\begin{figure}
\vbox to 3cm{\vfill\centerline{\fbox{Here is the figure 1}}\vfill}
\caption{Schematic equivalent circuit of the parallel array of Josephson-
junction. External current is injected from top to bottom. $R_S$ is a
small shunt resistance to deal with overdamped junctions.}
\label{fig1}
\end{figure}

We shall study first the array configuration. Let us consider a parallel array of
Josephson-junctions with alternating critical currents ($I_{c1}$ and
$I_{c2}$, generally $I_{c\alpha}$)
and alternating area plaquettes. The area differences cause  
alternating self-inductions $L_1$ and $L_2$. In order to deal with overdamped 
junctions, we shunt each of them with a small external resistance $R_s$ so
that we can neglect capacitance effects.
Figure 1 is a schematic picture of the circuit. We denote by $2\pi x_j$ the 
phase difference across the junction $j$. $I(t)$ is an external driving current. 
The equation for current conservation through the junction $j$  read as 
\begin{equation}
\frac{V_j}{R_s}\ + I_{c\alpha} \sin 2\pi x_j=I(t)+I_{j-1}^t-I_{j}^t= I(t)+
I_{j-1}^b -I_{j}^b
\label{eqcur}
\end {equation}
where 
\begin{equation}
V_j=\Phi_0 \dot{x}_j
\label{eqvol}
\end{equation}
being $\Phi_0$ is the flux quantum. The phase diference along a plaquette is 
given by 
\begin{equation}
x_{j+1} - x_j = \frac{\Phi_j}{\Phi_0}
\label{eqflux1}
\end{equation}
where
\begin{equation}
\Phi_j = \Phi^{ext} - L_{\alpha}(I_j^t + I_j^b)
\label{eqflux2}
\end{equation}
is the magnetic flux through a plaquette, sum of two contributions: 
external and induced magnetic fields\cite{Watanabe}. 
We set now $\Phi^{ext} = 0$.
Combining equations (\ref{eqcur}), (\ref{eqvol}), (\ref{eqflux1}) 
and (\ref{eqflux2}), and dividing the array in two sublattices, 
the equations of motion for the phases are:
\begin{equation}
\frac{\Phi_0}{R_s}\dot{x}_j + I_{c1}\sin 2\pi x_j = 
I(t) +\frac{\Phi_0}{2L_1}(x_{j-1} - x_j) + 
\frac{\Phi_0}{2L_2}(x_{j+1} - x_j)
\label{eqm1}
\end{equation}

\begin{equation}
\frac{\Phi_0}{R_s}\dot{x}_{j+1} + I_{c2}\sin 2\pi x_{j+1} = 
I(t) +\frac{\Phi_0}{2L_2}(x_{j} - x_{j+1}) +
\frac{\Phi_0}{2L_1}(x_{j+2} - x_{j+1})
\label{eqm2}
\end{equation}

Normalizing the equations by $I_L = \Phi_0 /2L_1$ and denoting by $K = 4 \pi
I_{c1} L_1/ \Phi_0$, $ \alpha= I_{c2}/I_{c1}$ and $\beta=L_1/L_2$ we obtain the
dimensionless equations of motion 

\begin{equation}
\dot{x}_j + \frac{K}{2\pi}\sin 2\pi x_j =
\tilde{I}(t) +(x_{j-1} - x_j) + 
\beta (x_{j+1} - x_j)
\label{eqmot1}
\end{equation}

\begin{equation}
\dot{x}_{j+1} + \alpha \frac{K}{2\pi} \sin 2\pi x_{j+1} =
\tilde{I}(t) + \beta (x_{j} - x_{j+1}) +
(x_{j+2} - x_{j+1}) 
\label{eqmot2}
\end{equation}
where time derivatives are given in the time scale $\tau = 2L_1/R_s$. Here
$K$ play the role of discreteness parameter 
\cite{Watanabe} and $\alpha$ and $\beta$ measure the asymmetry of the array.
Flux trapped in the array determines the boundary conditions,
therefore, if we have $n$ flux quanta in $N$ plaquettes, then
$x_{N+1} = x_1 + n$. Note that these equations of motion (\ref{eqmot1}) and (\ref{eqmot2}) correspond to the well known FK model in a
generalised asymmetric form. The overdamped dynamics for the symmetric model
have been reviewed in reference \cite{FloMa}. The mechanical
analogue of the JJ array we are dealing with is a set of pendula with two different lengths
coupled by two kinds of harmonic springs.

It is expected  for $K \rightarrow 0$ that the pinning potential for
fluxons vanishes. But we are interested in the opposite limit in which
fluxons are strongly pinned to the discrete lattice. Due to the spatial
asymmetry of the array we expect to have an asymmetric pinning potential. 

Now we will try to describe the fluxon as a single particle in the pinning 
periodic potential. Fluxon location is given by its center of mass (CM)
which can be expressed as \cite{Tsironis}
\begin{equation}
X_{CM}(x_1,\ldots ,x_N)= \frac{1}{2}+\sum_{i=1}^{N} i(x_{i+1} - x_i)
\label{CM}
\end{equation}
for any given phase configuration $\{x_i\}$. On the other hand, the
energy of the array is the sum of two contributions: Josephson and magnetic
energy. In the adimensional parameters defined above this energy reads as 
\begin{equation}
E(x_1, \ldots x_N)= \sum_{i=1,odd}^{N}\left [\frac{-K}{4\pi^2}(\cos 2\pi
x_i +\right . \alpha \cos 2\pi x_{i+1}) +
\left . \frac{\beta}{2} (x_{i+1}-x_i)^2 +\frac{1}{2}(x_{i+2}-x_{i+1})^2 \right ]
\label{energy}
\end{equation}
From these expressions we can define 
$E(X_{CM}) = min_{\{x_i\}} E(x_1 \ldots x_N)$ such 
that $\{x_i\}$ are kink configurations whose CM is $X_{CM}$. 

In order to calculate this potential, several approaches have been tried in
the context of non-linear discrete lattices. For low pinning (low $K$), a good
estimation is obtained using the collective coordinates method
\cite{Willis}. Essentially this method assumes a soliton profile for the
$\{x_i\}$
corresponding to the continuous limit (sine-Gordon) and, in some cases, an
analytical solution can be reached which only converges for
low values of $K$. We have adopted a more general way to (numerically) obtain
the potential profile $E(X_{CM})$. Maxima of this energy correspond
to saddle-points in the N-dimensional phase space. For configurations
containing one fluxon, N-1 directions are stable and one is unstable.
Such maxima points can be obtained using standard minimax methods. 
We have performed a linear stability
analysis around these points to get the direction of destabilization of
the maximum energy configuration. Using this saddle configuration as 
initial condition, we perturb it along the unstable direction
and study the relaxation (by numerical integration of equations 
(\ref{eqmot1}) and (\ref{eqmot2}), with $\tilde{I}(t)=0$ ) to the adjacent minima.
Then, the energy and CM along the trajectory
are computed  (according to equations (\ref{CM}) and
\ref{energy}). Using this method we have been able to 
calculate the energy profile
$E(X_{CM})$ for a trapped fluxon as a function of its CM.
Figure 2 shows potential profiles for a set of the model parameters
($K, \alpha, \beta$). In the following we set $K=4.0$. For $\alpha=0.5$ and
$\beta=1.0$ (different critical currents),
energy profile shows a double well structure symmetric respect to the bottom
of the wells. The $\alpha=1.0$, $\beta=0.5$ (different areas) potential is 
symmetric respect to
the tops. As expected, for any other values the potential profiles do
not show inversion symmetry. The values $\alpha=0.5$ and $\beta=0.5$ give
a good approximation to the asymmetric 
sawtooth potential used in the literature \cite{Magnasco}. 
It is interesting to note that, unlike other works in extented systems
\cite{Tsironis,Marchesoni}, neither the on-site potential nor the
interparticle potential are asymmetric in the field variables.

\begin{figure}
\vbox to 6cm{\vfill\centerline{\fbox{Here is the figure 2}}\vfill}
\caption[]{Energy profile $E(X_{CM})$ for different values
of asymmetry parameters $(\alpha, \beta)$ (a) $\alpha=0.5, \beta=1.0 $. Same
plaquette areas and different critical currents. (b) $\alpha=1, \beta=0.5 $.
Different plaquettes and the same critical currents. 
(c) $\alpha=0.5, \beta=0.5 $ gives ``ratchetlike'' potential.}
\label{fig2}
\end{figure}

We will study now the dynamical behavior of a fluxon in the asymmetric
lattice. For all simulations, we take $N=30, n=1, K=4.0, \alpha=0.5$ and
$\beta=0.5$. We drive the system out of thermal equilibrium by applying an
external ac bias currents.
The $\tilde{I}(t)$ term in equations (\ref{eqmot1}) and (\ref{eqmot2}) 
is then expressed as
\begin{equation}
\tilde{I}(t) = \tilde{I}_{ac} \sin \omega t + \xi (t).
\label{driving}
\end{equation}
Here, $\xi(t)$ is  white noise ($\langle \xi(t) \rangle=0$ and $\langle
\xi(t)\xi(t') \rangle = 2 D \delta(t-t')$)\footnote{Where
$D=\frac{k_BT}{\Phi^2_0/2L_1}$ is the temperature in units of the magnetic
energy.}
which, in absence of other forces, brings the system to thermal equilibrium.
Thus, the equations of motion take the form of a system of stochastic
differential equations. We have solved them using a fourth order Runge-Kutta
method for the deterministic part and third order for the stochastic one
\cite{GH}.

We will concentrate first on the deterministic ($T=0$) dynamics.
We check the ratchet behavior measuring the critical current of the array
under dc driving. For positive
driving we find a fluxon deppining current $I_{dp}^+ \approx 0.18$ whereas
for negative $I_{dp}^- \approx 0.31$ (for the symmetric case, $\alpha=1.0$
and $\beta=1.0$, $I_{dp} \approx 0.19$). Under ac currents we should observe
a IV curve showing rectification of the external current and
the maximum efficienty for rectification (in the low frequency limit) of the
ac currents is expected to be around $I_{dp}^+$ \cite{Astumian}.
Figure 3 shows the IV (voltage versus amplitude $I_{ac}$) curves for
$I_{dc}=0$ and $T=0$ for different
frequencies $\omega$. Low frequency response clearly resembles to that found
for a single particle \cite{Hangi}. As it was noted in one particle
simulations, voltage is quantised:
\begin{equation}
V  = \frac{\Phi_0}{\tau} \sum_{i=1}^N \langle \dot{x}_i \rangle 
= \frac{2 \Phi_0}{\tau} (\frac{p}{q} \omega),
\label{voltaje}
\end{equation}
being p and q integer numbers \cite{note2}. For finer resolution, 
the IV curves appears
to have a devil's staircase structure \cite{Hangi} (see inset of figure 3c). 
This quantisation result can be straightforward obtained 
by assuming that, in the
overdamped ac dynamics, the phases $x_i(t)$ can be expressed by a two
dimensional envelope function \cite{FloMa,FloFa}
$x_i(t) = \zeta(i- V t/\Phi_0,\omega t)$ \cite{note}.
Now, imposing energy balance along the trajectory   
\begin{equation}
\sum_{i=1}^{N} \int_{0}^{\infty} \dot{x}_i^2 dt = \sum_{i=1}^{N}
\int_{0}^{\infty}
\dot{x}_i \tilde{I}(t) dt
\label{balener}
\end{equation}
we find that only voltages commensurate with the current frequency
give non-zero
contribution to the right side of eq. \ref{balener}. 

\begin{figure}
\vbox to 6cm{\vfill\centerline{\fbox{Here is the figure 3}}\vfill}
\caption[]{Voltage-Current (IV) curves at $T=0$ and ac current, for different
frequencies $\omega$ in units of $1/\tau$. (a) $\omega= 2\pi0.01$ . 
(b)  $\omega= 2\pi0.05$. (c) $\omega=2\pi0.1$.}  
\label{fig3}
\end{figure}
\begin{figure}
\vbox to 5cm{\vfill\centerline{\fbox{Here is the figure 4}}\vfill}
\caption[]{Voltage-Current (IV) curves for ac current at $D=0.001$
and $\omega= 2\pi0.01$. Thick line shows the comparison with
equation 3 of reference \cite{Magnasco}.}
\label{fig4}
\end{figure}

Quantised steps appears rounded at non-zero temperature as a
consequence of multiples hops between dynamically close
attractors, corresponding to the mode-locking trajectories\cite{rounding}. 
In the
adiabatic (low frequency) and low temperature limit we have fitted the
IV curves with the solution of Magnasco \cite{Magnasco} for a sawtooth
ratchet potential. We have found a good fit (see figure \ref{fig4})
for asymmetry ratio  $\lambda_1
/\lambda_2 = 0.56$ and energy barrier $\Delta E=0.16$  obtained from 
the computed energy profile (see figure 2c). {At low frequencies the
steps disappear, even for very low temperatures. At high frequency} the
steps seem to be more stable against thermal fluctuations
and probably could be observed in real experiments.

In summary, we have proposed a new experiment for DM based in the motion of a
fluxon in a one-dimensional Josephson-junction array. We have calculated the 
effective pinning potential for the fluxon and found the parameters needed
to have almost a sawtooth one. The simulated IV curves mimic the behavior of
a single particle in an asymmetric periodic potential.  

The actual feasibility to fabricate these kinds of arrays along with the
simplicity of our design make this experiment really accesible \cite{note3}.
Evenmore,
the possibility of introducing in the array a controlled number of fluxons
could serve to study the influence of interaction in  directional motion 
brownian particles \cite{Vicsek}. Preliminary results in this directions show
that complex behavior dominates the multifluxon dynamics.

\stars
We acknowlege J.C. Ciria and E. Tr\'{\i}as for discussions and L.M. Flor\'{\i}a 
for many useful suggestions and critical reading of this work. 
JJM is supported by a Fulbright/MEC fellowship. 
Finantial support is acknowledged to DGES (PB95-0797), Spain.


\begin{thebibliography}{99}
\bibitem{note2} The factor 2 of Eq. (\ref{voltaje}) appears as consequence
of the double periodicity of the pinning potential. Integer number $p$ and $q$
can be related with the spatial and temporal periodicity of $x_i$. See ref
\cite{FloMa,FloFa} for more details.
\bibitem{note} The envelope function $\zeta(x,y)$ is periodic in the second
variable $\zeta(x,y)= \zeta(x,y+2\pi)$ and  $\zeta(x,y_o)$ has the form of a
kink.
\end{thebibliography}
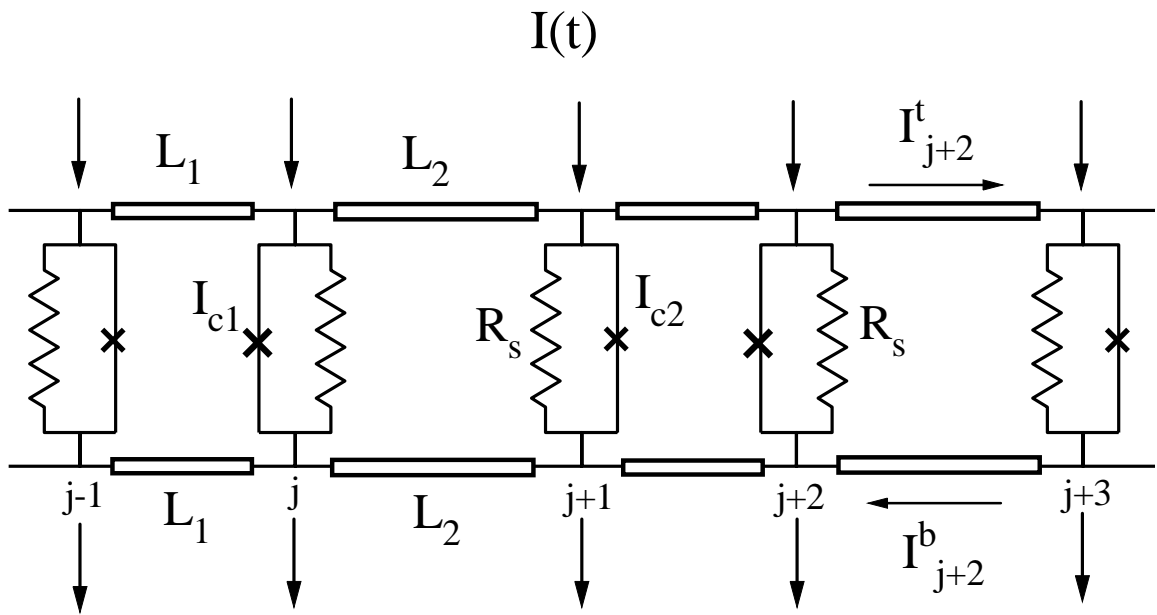
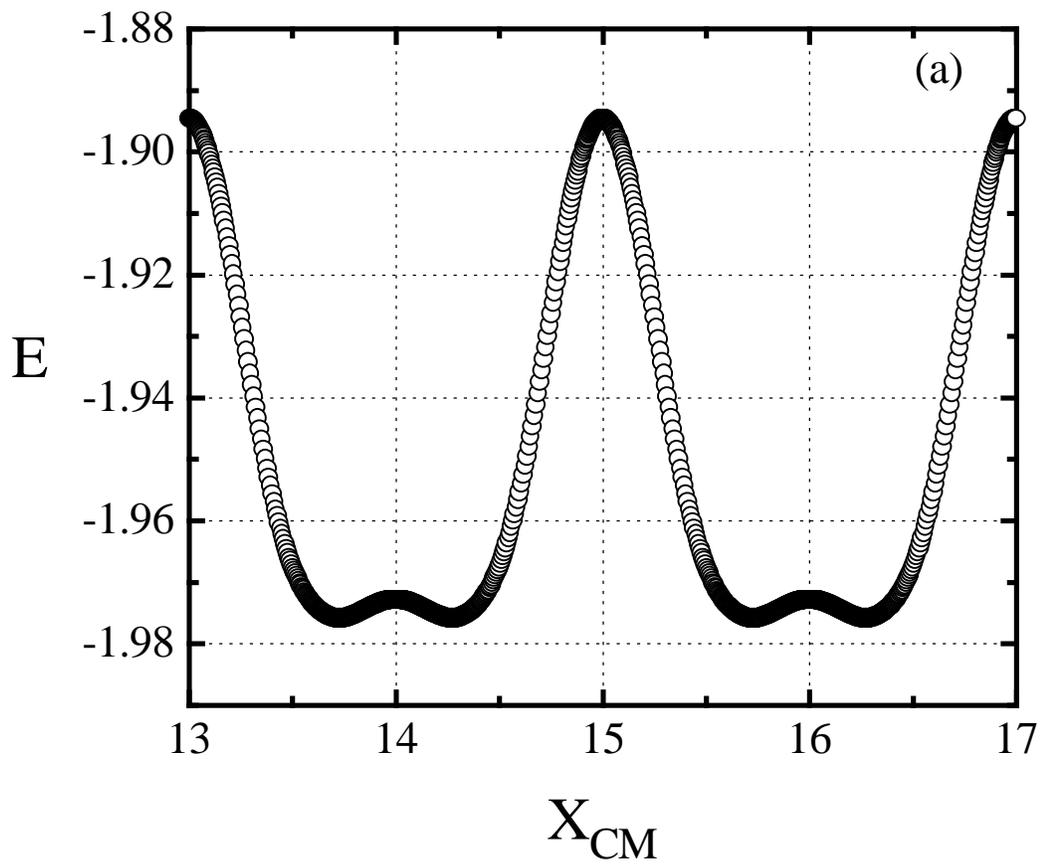
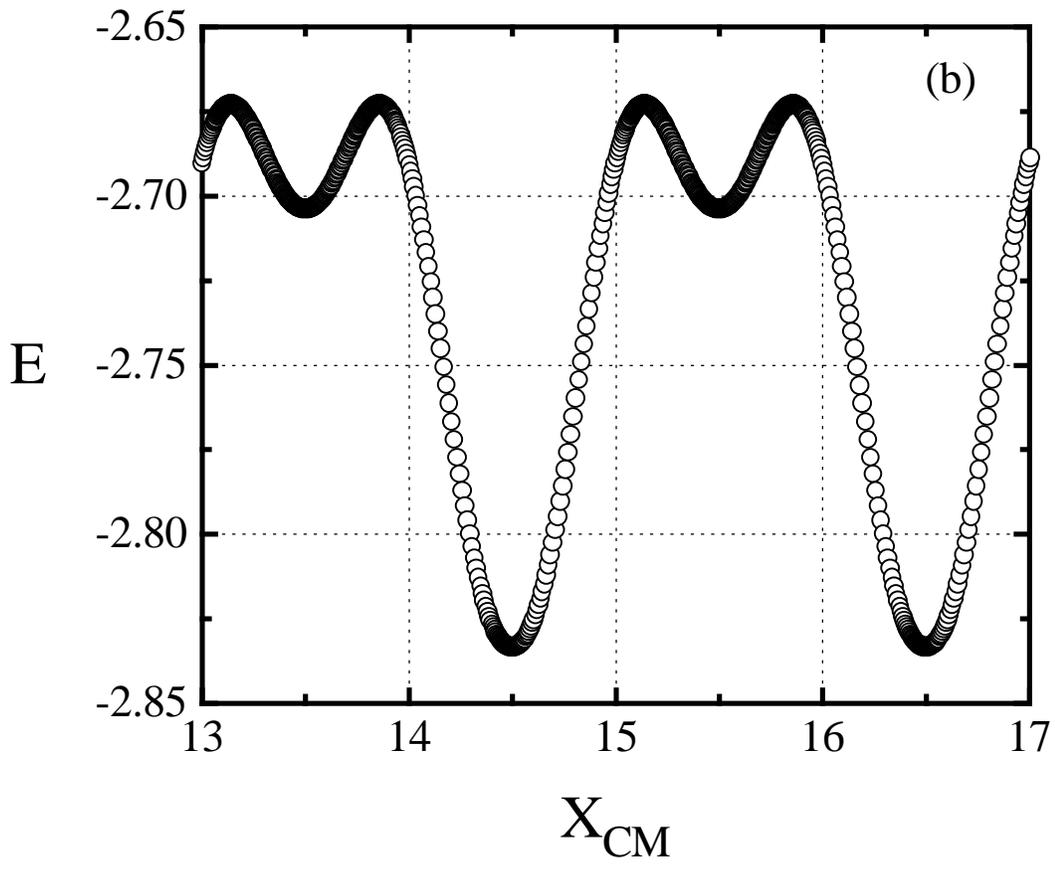
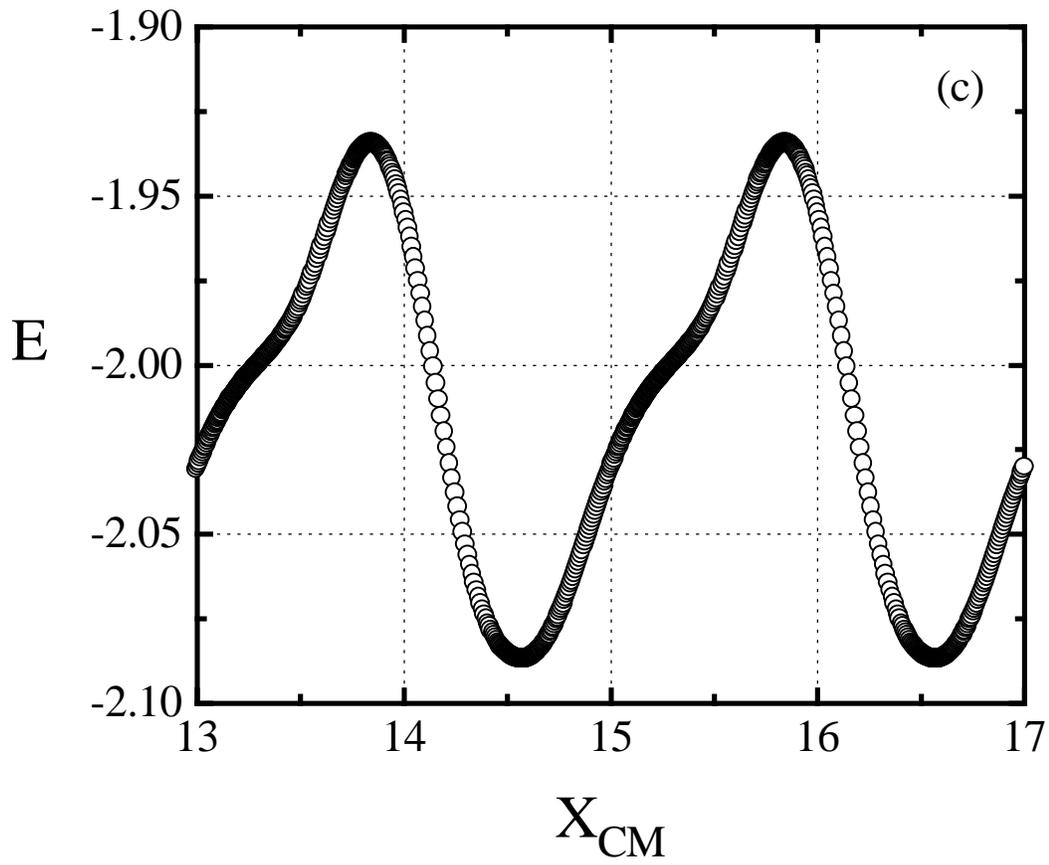
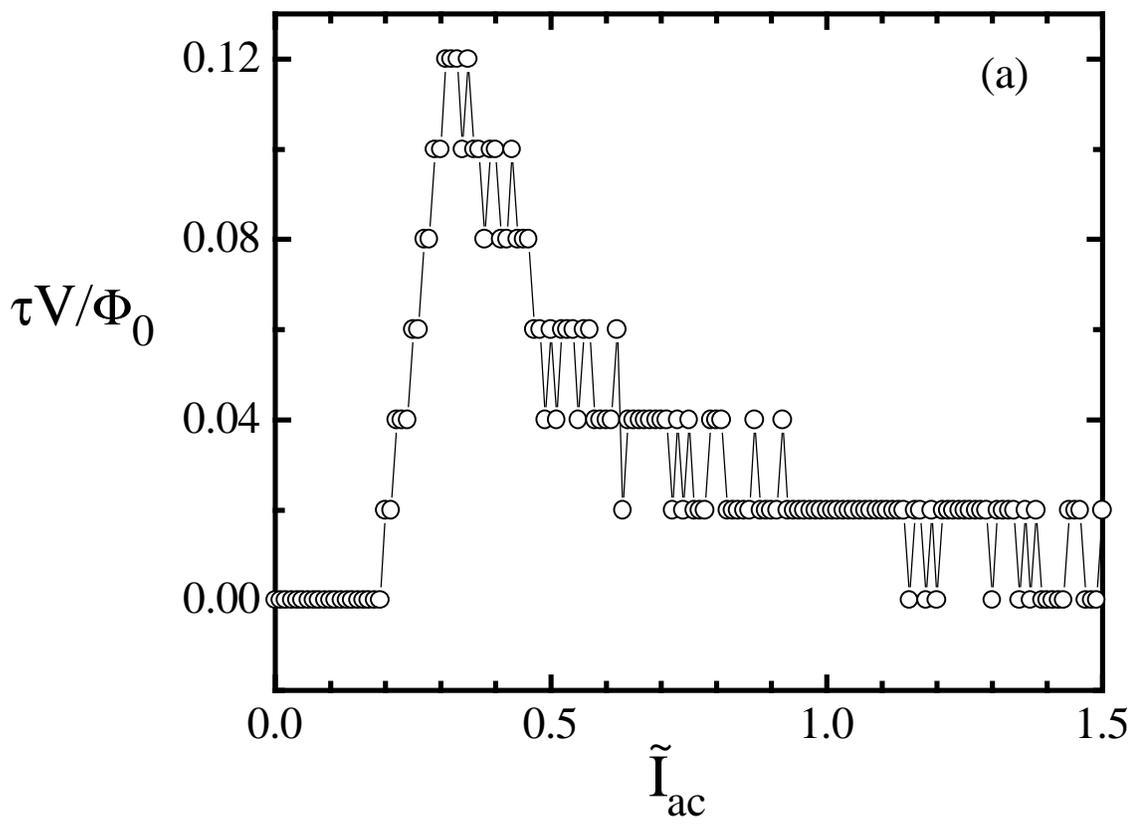
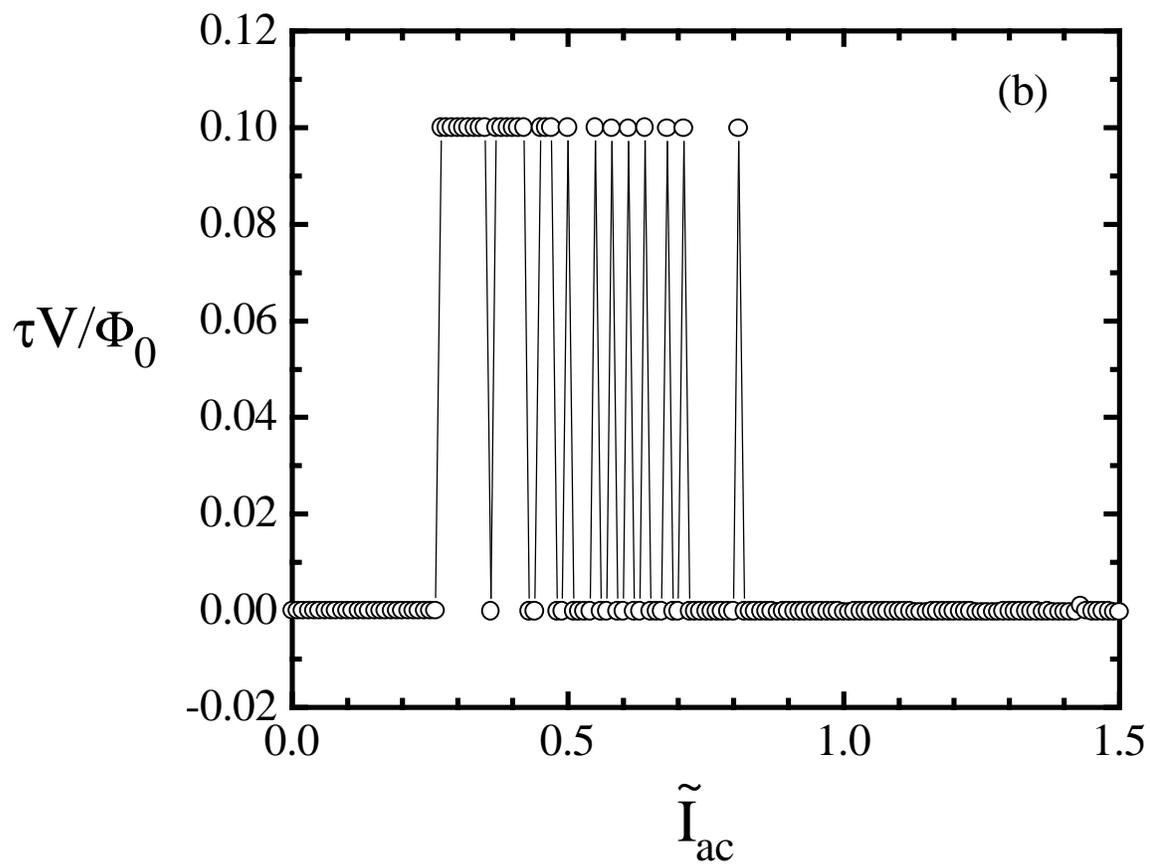
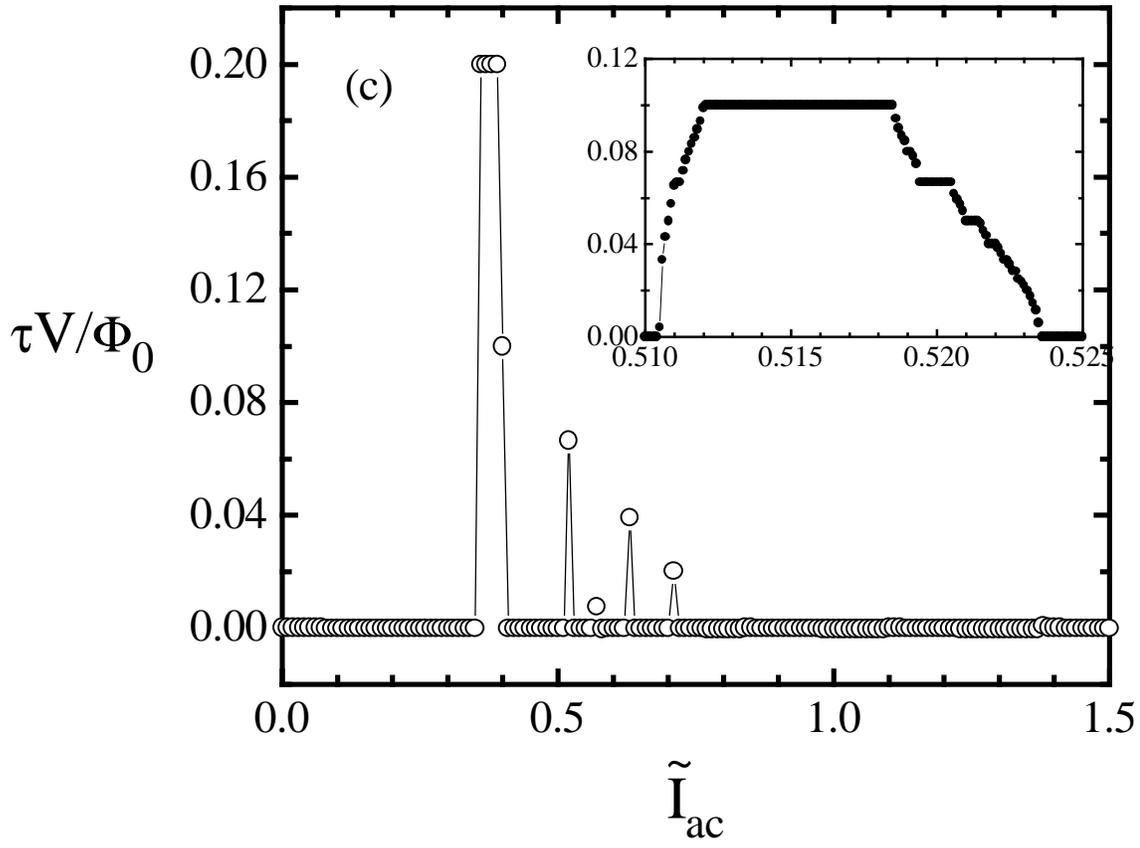
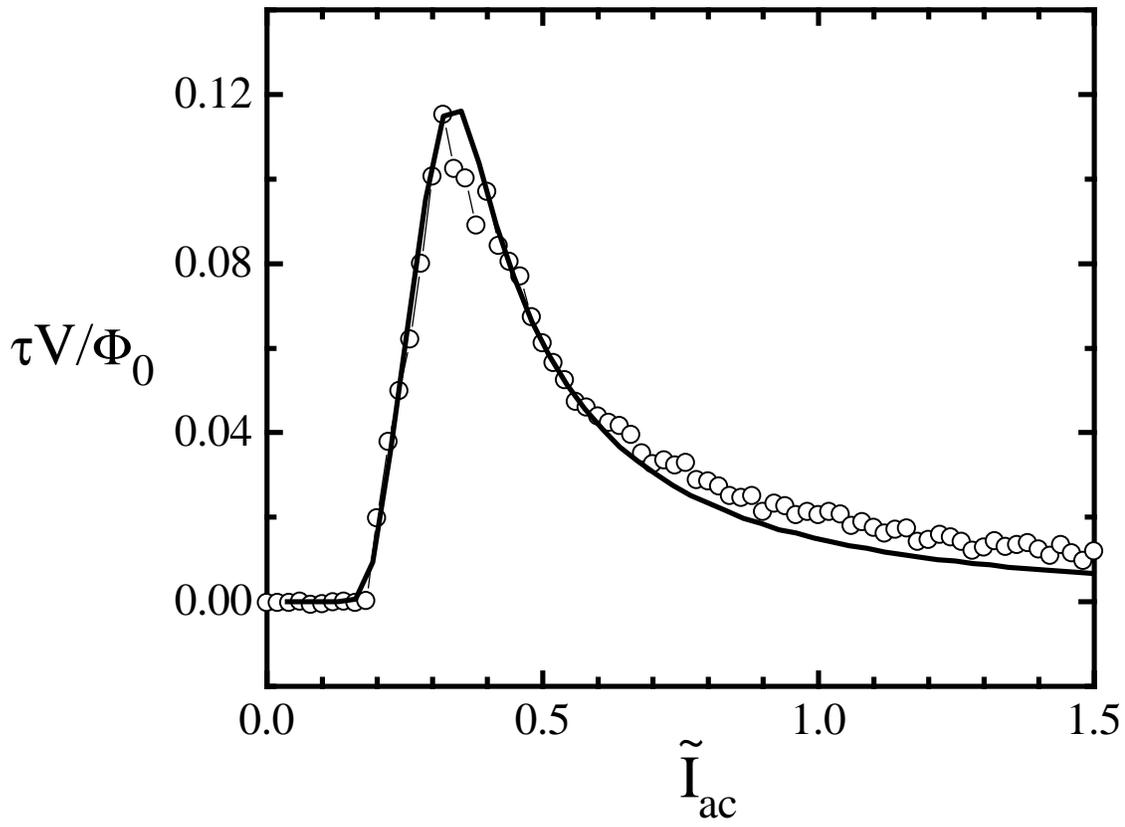
\end{document}